\documentclass{article}

\usepackage{PRIMEarxiv}

\usepackage[utf8]{inputenc} 
\usepackage[T1]{fontenc}    
\usepackage{hyperref}       
\usepackage{url}            
\usepackage{booktabs}       
\usepackage{amsfonts}       
\usepackage{nicefrac}       
\usepackage{microtype}      
\usepackage{lipsum}
\usepackage{fancyhdr}       
\usepackage{graphicx}       
\graphicspath{{media/}}     
\usepackage{tabularx}
\usepackage{multirow}
\usepackage{makecell}
\usepackage{caption}
\usepackage{subcaption}
\usepackage{multirow}
\usepackage{colortbl}
\usepackage{float}
\usepackage{amsmath}
\title{Pairwise Ranking Loss for Multi-Task Learning in Recommender Systems
}

\author{
 Furkan Durmus \thanks{These authors contributed equally to this work.} \\
  AI Enablement\\
  Huawei Türkiye R\&D Center\\
  Istanbul, Turkey \\
  \texttt{furkan.durmus2@huawei.com} \\
   \And
 Hasan Saribas \footnotemark[1] \\
  AI Enablement\\
  Huawei Türkiye R\&D Center\\
  Istanbul, Turkey \\
  \texttt{hasan.saribas2@huawei-partners.com}
  \And
 Said Aldemir \\
  AI Enablement\\
  Huawei Türkiye R\&D Center\\
  Istanbul, Turkey \\
  \texttt{said.aldemir1@huawei.com}
  \And
 Junyan Yang \\
  Huawei Technologies Co Ltd\\
  Shenzhen, China \\
  \texttt{yangjunyan4@huawei.com}
  \And
 Hakan Cevikalp \\
 Electrical and Electronics Engineering Department \\
  Eskisehir Osmangazi University\\
  Eskisehir, Turkey \\
  \texttt{hakan.cevikalp@gmail.com}
}

\begin{document}

\maketitle

\begin{abstract}
Multi-Task Learning (MTL) plays a crucial role in real-world advertising applications such as recommender systems, aiming to achieve robust representations while minimizing resource consumption. MTL endeavors to simultaneously optimize multiple tasks to construct a unified model serving diverse objectives. In online advertising systems, tasks like Click-Through Rate (CTR) and Conversion Rate (CVR) are often treated as MTL problems concurrently. However, it has been overlooked that a conversion ($y_{cvr}=1$) necessitates a preceding click ($y_{ctr}=1$). In other words, while certain CTR tasks are associated with corresponding conversions, others lack such associations. Moreover, the likelihood of noise is significantly higher in CTR tasks where conversions do not occur compared to those where they do, and existing methods lack the ability to differentiate between these two scenarios. In this study, exposure labels corresponding to conversions are regarded as definitive indicators, and a novel task-specific loss is introduced by calculating a \textbf{p}air\textbf{wise} \textbf{r}anking (PWiseR) loss between model predictions, manifesting as pairwise ranking loss, to encourage the model to rely more on them. To demonstrate the effect of the proposed loss function, experiments were conducted on different MTL and Single-Task Learning (STL) models using four distinct public MTL datasets, namely Alibaba FR, NL, US, and CCP, along with a proprietary industrial dataset. The results indicate that our proposed loss function outperforms the BCE loss function in most cases in terms of the AUC metric.
\end{abstract}


\section{Introduction}

With the rapid growth of online recommender systems, click-through rate (CTR) prediction is no longer the only main focus in industrial applications. Increasingly, various online metrics, such as different types of conversion rates, are also being optimized to meet user needs and the platform’s profit targets. For example, online advertising recommender systems aim to expose apps to users who are not only likely to click on them but also to download, activate, and make purchases within these apps \cite{cvrattention, ctnocvr}. Effective Cost Per Mille (eCPM) is used to evaluate the value of an advertisement display, and it is calculated by several factors, including the predictions of both CTR and CVR \cite{star, 3MN}. Therefore, it is quite useful and natural to apply MTL to recommender systems to optimize different metrics simultaneously.

MTL aims to optimize multiple tasks by building a single model for different objectives, which can improve the performance by utilizing the correlation between tasks and greatly reduce online consumption by using one model to serve all different task predictions \cite{caruana1997multitask, esmm, mmoe, ple}. Prior works mainly focus on the new structure in the network, investigating how to model the complex relationship including the similarity as well as the conflicts between the tasks \cite{hinet, causalint}. However, a causal connection sometimes occurs between these tasks. For example, in advertising systems, the CTR and CVR tasks are interdependent; a conversion cannot occur without a preceding click. Using an e-commerce recommender system as an example of the overall cycle of CTR and CVR estimation, the platform first recommends various products to users. The users then click on some of these products and ultimately purchase their preferred ones. This intrinsic and sequential user action pattern — impression -> click -> conversion — can be used to model both CTR and CVR \cite{esmm, ctnocvr}. In online advertising and recommender systems, the predicted Click \& Conversion Rate (pCTCVR) is calculated as the product of the predicted Click-Through Rate (pCTR) and the predicted Conversion Rate given a click (pCVR) \cite{ip, escd}. This relationship, expressed as \( \text{pCTCVR} = \text{pCTR} \times \text{pCVR} \), provides a comprehensive measure of an ad's effectiveness by considering both the likelihood of a user clicking on the ad (pCTR) and the likelihood of converting after the click (pCVR). By combining these probabilities, pCTCVR helps optimize ad selection and ranking, prioritizing ads that are more likely to result in conversions, thereby enhancing revenue generation and user experience on the platform. From the perspective of eCPM, merely obtaining a click is insufficient; the ultimate goal is to achieve a conversion to generate profits. Therefore, there exists a mutual relationship between these two types of tasks, which traditional network structures often struggle to model explicitly.

\begin{figure}[]
  \centering
  \includegraphics[width=0.7\linewidth]{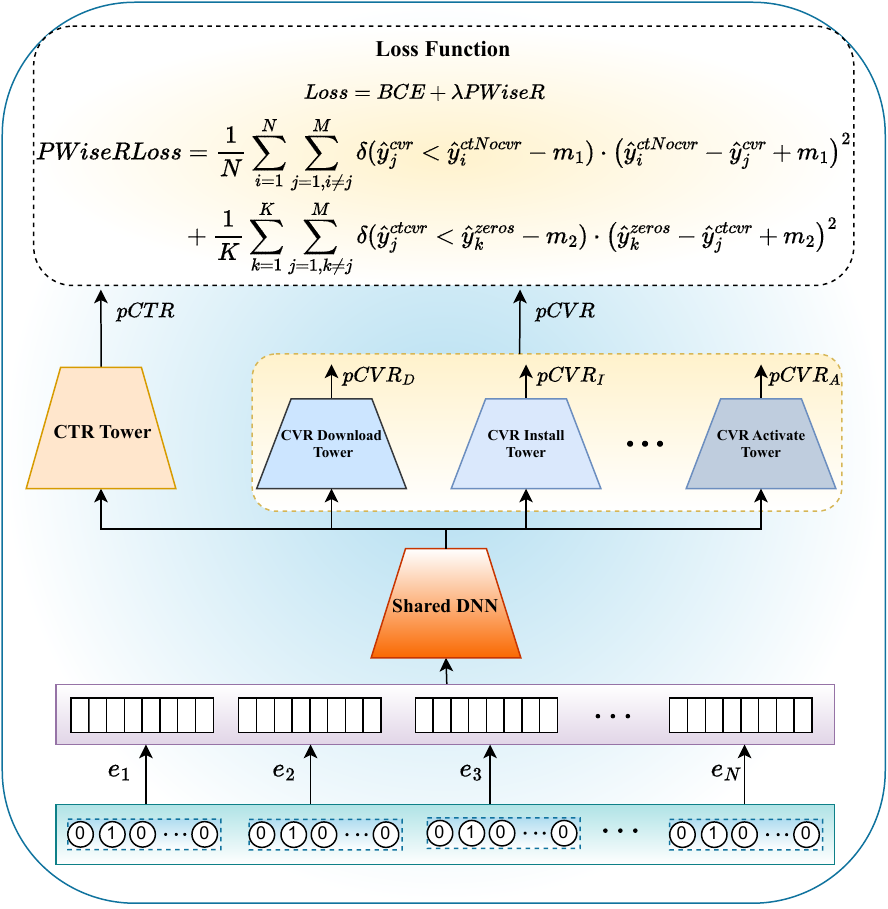}
  \caption{Demonstration of the proposed loss function for the training phase together with the overall MTL architecture.}
  \label{Architecture}
\end{figure}

Let’s suppose that we have two training samples, both with positive click labels: the first one with a positive conversion label and the second one with a negative conversion label. In an actual industrial situation, it is highly believed that the first sample will have a higher eCPM, along with more reliable click and conversion probabilities. The model needs to rank the first sample higher than any others without any conversion. Moreover, the likelihood of a sample being noise is much higher in the CTR task when conversions do not occur, compared to cases where they do. 

Normally, it is quite common for an MTL model to be trained with these two samples using binary cross-entropy (BCE) loss, which helps the model to learn and predict the task’s ranking score in a well-calibrated manner \cite{ple, mmoe, esmm}. However, it cannot distinguish the above case, especially for the ranking part.

Both CTR and CVR models are utilized in determining this ranking. Since tasks resulting in conversions yield higher revenue, hence a higher eCPM metric, instances where the CVR value is 1 during the training of CTR models imply that these models are more sensitive to such instances, indicating they will return more compatible rankings with CVR models.

In this paper, a novel task-specific pairwise ranking loss is proposed for the above problems. This loss function, used in conjunction with BCE, generates a pairwise loss between each instance where a conversion occurs and those where it does not. This enables each candidate's items to be ranked more precisely. The proposed PWiseR loss ensures that CTR models give more weight to exposures where conversions occur, resulting in a more robust model against noise. Additionally, since CTR models also utilize CVR labels, these two models will work more harmoniously during the advertising ranking process. This approach prioritizes tasks where conversions occur, recommending higher revenue-generating items.

The primary contributions of this paper can be outlined as follows:
\begin{itemize}
\item The proposed loss function ensures higher scores are assigned to instances where conversions occur, thereby facilitating the network's avoidance of noisy examples and improving the advertisement candidate ranking process.
\item The proposed approach can be applied not only to MTL but also to STL, provided that both CTR and CVR labels are available, and it is designed to be model-agnostic.
\item In order to demonstrate the effectiveness of the proposed method, results are shown for both 3 different MTL methods and STL methods on 4 different MTL public datasets, and experiments are also performed on an industrial dataset and the results show superiority of PWiseR loss. 
\end{itemize}

\section{Related Works}
MTL has gained attraction in the field of recommender systems due to its ability to leverage task relationships, shared knowledge to improve recommendation quality, and optimize multiple tasks simultaneously \cite{caruana1997multitask, ruder2017overview, xi2021modeling}.

The shared-bottom model, which is widely used in MTL, was proposed by Caruna \cite{caruana1997multitask,caruana1993multitask}. This structure uses shared layers to extract low-level features from input data, that is common across multiple tasks. Covington et al. \cite{covington2016deep} used shared-bottom structure for video recommendation. Although this design is successful in reducing over-fitting, it can face optimization challenges stemming from task discrepancies, as all tasks are required to employ identical parameters in the shared-bottom layers. 
To address this, MMOE \cite{mmoe} employs task-specific gates, an extended version of the mixture of experts \cite{moe}, allowing different fusion weights in multi-task learning. PLE \cite{ple} further refines this by distinguishing between shared and task-specific components, using a progressive routing mechanism to separate deeper semantic knowledge. This enhances the efficiency of joint representation learning and information routing across tasks.

An end-to-end framework called ESDF \cite{esdf} is proposed to address the delayed feedback problem by introducing a time delay model that analyzes the expected time duration until the user events as click and conversion happen. 
Furthermore, Tan et al. proposed a unified ranking model \cite{mtms} for a multi-task and multi-scene online advertising scenario that consists of prediction of CTR and CVR in multiple services for users such as news feed, search engines and product suggestions. 
There are studies that proposed custom loss functions to enhance modeling of CTR and CVR \cite{mmn, mebs, dupn}. Zhang et al. proposed MMN \cite{mmn}  with a dynamically weighted loss that is computed within each mini-batch in order to address the loss scale imbalance issue for multi-domain and multi-type CVR prediction problem. DUPN \cite{dupn} is a unified architecture that learns a general and universal user representation from multiple e-commerce tasks. The structure combines recurrent neural networks, attention and MTL concept. Same user representation is shared by each block in the network architecture with specific loss functions to learn weights of ranking features to maximize the CTR and CTCVR prediction. 

Gong et al. \cite{mebs} introduced bid shading for bid price adjustment in multi-slot advertisement by using a multi-task framework. The proposed MEBS method consists of win rate model that predicts whether the auction is won with a specific bid price, pCTR calibration model for ad position-aware learning, and shading ratio model for predicting optimal shading ratio and expected surplus. These models in MEBS framework are trained by using shared embeddings with cross-entropy loss or two custom losses that are expected surplus loss and shading ratio loss functions. However, these studies could not address distinguishing the samples with CVR label of 1 from the noisy samples and assigning higher scores for them in the ranking stage with a specific loss function.

\section{Method}
\subsection{Problem Definition}
Let us consider the dataset \( S = \{x_i, y_{ctr}, y_{cvr}\} \mid _{i=1} ^{n} \), where \( n \) is the number of samples in the dataset. Each \( x_i \) represents an individual sample that includes various fields such as user information, item details, and combined features. The labels \( y_{ctr} \) and \( y_{cvr} \) are binary indicators for the \( i \)-th sample, representing whether there was a click (\( y_{ctr} \)) and whether there was a conversion (\( y_{cvr} \)).

\begin{table}[]
    \centering
\caption{Demonstration of all possible scenarios with relevant labels in the MTL system.}
\resizebox{0.5\textwidth}{!}{
    \centering
    \begin{tabular}{cccc} \hline  
         \textbf{Scenarios}&  \pmb{$y_{ctr}$}&  \pmb{$y_{cvr}$}&  \textbf{Comments}\\ \hline  
         1&  0&  0&  no click and no conversion\\ \hline  
         2&  1&  0&  click without conversion\\ \hline  
         3&  1&  1&  click with conversion\\ \hline 
    \end{tabular}}
    \label{tab:my_label}
\end{table}

Table \ref{tab:my_label} summarizes the possible outcomes in a MTL system concerning two binary labels: \( y_{ctr} \) (click-through rate) and \( y_{cvr} \) (conversion rate). The table defines three scenarios: scenario 1, where \( y_{ctr} = 0 \) and \( y_{cvr} = 0 \), indicating no click and no conversion; scenario 2, where \( y_{ctr} = 1 \) and \( y_{cvr} = 0 \), representing a click without a conversion; and scenario 3, where \( y_{ctr} = 1 \) and \( y_{cvr} = 1 \), signifying a click with a conversion. 
In the literature, data from all these scenarios are typically utilized for CTR prediction. However, considering user interactions, scenario 2 is more likely to contain noise compared to scenario 3. For instance, scenario 2 may include various erroneous clicks, such as those from bot traffic, accidental user clicks, click fraud, or click injections. In contrast, scenario 3 involves a conversion, rendering it a much more reliable example. It is imperative to communicate this distinction to the model in the most appropriate manner to enhance predictive accuracy.
Additionally, it is evident that examples from scenario 3, which involve conversions, have higher bidding values and consequently generate much higher returns compared to examples from scenario 2.

For the reasons summarized above, we propose a loss function that identifies the examples from scenario 3 as valuable instances and enforces the classifier model by giving higher weights to these important examples during training.

\subsection{Proposed Loss Function}

Figure \ref{Architecture} illustrates the general MTL architecture during the training phase, incorporating the proposed loss function. This loss function is used in conjunction with Binary Cross Entropy (BCE), which is commonly employed in recommender systems \cite{deepfm, mmbattn, MaskNet, stec}, and can be expressed as follows:

\begin{equation}
Loss = BCE +\lambda PWiseR
\end{equation}
where $\lambda$ is a hyperparameter, often called the balancing term, that controls the importance of the second term, PWiseR loss, in the total loss function. As in most other methods, BCE loss is used to distinguish between instances belonging to labels 0 and 1, while pairwise ranking loss is proposed to enforce the network to learn the difference between instances with and without conversion. Our proposed pairwise ranking loss can be written as,
{\footnotesize
\begin{equation}
\begin{aligned}
PWiseR &= \frac{1}{N} \sum_{i=1}^{N} \sum_{j=1,i\neq j}^{M}\delta(\hat {y}^{cvr}_j < \hat{y}^{ctNocvr}_i -  m_{1}) \cdot \big(\hat{y}^{ctNocvr}_i -\hat {y}^{cvr}_j + m_{1}\big)^2 \\
&+ \frac{1}{K} \sum_{k=1}^{K} \sum_{j=1,k\neq j}^{M}\delta(\hat {y}^{cvr}_j < \hat{y}^{zeros}_k - m_{2}) \cdot \big(\hat{y}^{zeros}_k -\hat {y}^{cvr}_j + m_{2})^2
\end{aligned}
\end{equation}}
where, $N$, $K$ and $M$ respectively represent the number of samples with labels $y_{ctr} = 1 \& y_{cvr}=0$ (only clicked no conversion), ${y}_{cvr} = 1$ (conversion occurred) and ${y}_{ctr} = 0$. The terms $\hat{y}^{ctNocvr}$, $\hat{y}^{cvr}$ and $\hat{y}^{zeros}$ $\in (0,1)$ are the prediction scores for the same ordered labels. $\delta(condition)$ is a delta function that returns 1 if the condition is true and 0 otherwise. The constants $m_1$ and $m_2$ are selected margin parameters.

The provided loss function, PWiseR loss, consists of two terms, the first term penalizes instances where $\hat{y}^{cvr}$ is less than $\hat{y}^{ctNocvr}$ by at least $m_1$ with the penalty being to the square of ($\hat{y}^{ctNocvr}_i -\hat {y}^{cvr}_j + m_{1}$). This term will allow the network to have more confidence in tasks that result in conversion compared to those that only result in clicks, and it will tend to assign higher prediction scores to these examples. The second term, similar to the first term, will penalize cases where $\hat{y}^{cvr}$ is less than $\hat{y}^{zeros}$ by at least $m_2$ with the penalty being to the square of ($\hat{y}^{zeros}_i -\hat {y}^{cvr}_j + m_{2}$). With this term, in addition to assigning higher scores to tasks resulting in conversion, it ensures that these examples are more distinguishable from tasks where no clicks occur.

\section{Experiments}
\subsection{Experimental setup}
This section explains the experiments carried out to prove the effectiveness of PWiseR loss. For reproducibility reasons, 4 different publicly available Alibaba datasets were used that contain CTR and CVR labels. Also in-house private industrial dataset was used in addition to the public datasets. Table \ref{ctr_ratios} shows the number of impressions and how much of it consists of clicks and conversions. Alibaba-CCP dataset was obtained from traffic logs on Taobao \cite{ma2018entire}, Alibaba-NL, Alibaba-FR and Alibaba-US datasets were obtained from traffic logs of the search system on the Aliexpress platform for different countries \cite{peng2020improving}. The train-test split of the Ali-CPP dataset was carried out as in \cite{xi2021modeling}, and the splits for the FR, US and NL datasets were carried out as in \cite{zhu2021learning}. FuxiCTR 2.1.3 \cite{ZhuLYZH21, ZhuDSMLCXZ22} framework was used for reproducibility purposes.

\begin{table}[H]
\centering
\caption{The table shows the number of impressions, clicks and conversions for all public datasets.}
\label{ctr_ratios}
\begin{tabular}{lcccc}
\textbf{Alibaba} & \multicolumn{1}{c}{\textbf{FR}} & \multicolumn{1}{c}{\textbf{NL}} & \multicolumn{1}{c}{\textbf{US}} & \multicolumn{1}{c}{\textbf{CCP}} \\ \hline
Impression       & 27035601                        & 17717195                        & 27392613                        & 85316519                         \\
Click            & 542753                          & 381078                          & 449608                          & 3317703                          \\
Conversion       & 14430                           & 13815                           & 10830                           & 17167                            \\ \hline
CTR              & 2.01\%                          & 2.15\%                          & 1.64\%                          & 3.89\%                           \\
CVR              & 2.66\%                          & 3.63\%                          & 2.41\%                          & 0.52\%                           \\
CTCVR            & 0.05\%                          & 0.08\%                          & 0.04\%                          & 0.02\%                           \\ \hline
\end{tabular}
\end{table}

Since our proposed loss function works effectively in both MTL and STL scenarios, experiments were conducted for both scenarios. Shared Bottom, MMoE and PLE models were used as MTL models. MaskNet \cite{MaskNet} and DNN models were used for STL experiments. To make fair comparisons, the model parameters in multitasking were chosen as the parameters in \cite{zhu2021learning}, and the same parameters were used in both three models and two losses. The parameters of MaskNet and DNN in the STL experiments were set as default parameters in FuxiCTR. However, unlike BCE, the PWiseR loss utilizes parameters $\lambda$, $m_1$, and $m_2$. Hyperparameter tuning was performed solely for these parameters using the grid search method, and the obtained parameters are presented in Table \ref{hyper} for Alibaba-US dataset.

\begin{table}[H]
\centering
\caption{The table shows optimum parameters for Alibaba-US dataset results of PLE model that uses PWiseR loss.}
\resizebox{0.7\textwidth}{!}{
\begin{tabular}{lcccccccc}
\hline
\textbf{H. P.} & \pmb{$m1$} & \pmb{$m2$} & \pmb{$\lambda$} & \textbf{lr} & \textbf{BS} & \textbf{emb. dim} & \textbf{\#expert} & \textbf{w. decay} \\ \hline
\textbf{Value}           & 0.3         & 0.3         & 0.1           & 1e-3                    & 2048                 & 128                     & 8                    & 1e-6                  \\ \hline
\end{tabular}}
\label{hyper}
\end{table}

\textbf{Performance on MTL models}:
Table \ref{table_results_public} shows a detailed comparison of BCE and the proposed PWiseR losses on common MTL models, including MMoE, PLE, and SharedBottom, for CTR and CTCVR tasks across four different public datasets from Alibaba.

The proposed loss function has shown better performance than BCE in the Alibaba FR, NL, and CCP datasets, except for the SharedBottom model on the Alibaba FR dataset and the MMoE model on the Alibaba CCP dataset for the CTCVR case. In Alibaba US, MMoE demonstrates improved CTCVR with PWiseR, while BCE yields better results for CTR. Similarly, PLE marginally enhances CTR with PWiseR, while BCE performs better for CTCVR. SharedBottom benefits from PWiseR for CTR but BCE outperforms for CTCVR. Overall, PWiseR generally enhances CTR and CTCVR across most datasets and models, although exceptions exist where BCE performs better.




\begin{table*}[t]
\centering
\caption{The table compares PWiseR loss performance on Alibaba's public datasets. AUC metric was used for comparison and results shown in bold are the best results for each method's corresponding loss function.}
\resizebox{0.95\textwidth}{!}{
\label{table_results_public}
\begin{tabular}{lllllllllll}
\multicolumn{1}{c}{\multirow{2}{*}{\textbf{Alibaba US}}} & \multicolumn{2}{c}{\textbf{CTR}}                                       & \multicolumn{2}{c}{\textbf{CTCVR}}                                     &  & \multicolumn{1}{c}{\multirow{2}{*}{\textbf{Alibaba FR}}}  & \multicolumn{2}{c}{\textbf{CTR}}                                       & \multicolumn{2}{c}{\textbf{CTCVR}}                                     \\
\multicolumn{1}{c}{}                                     & \multicolumn{1}{c}{\textbf{BCE}} & \multicolumn{1}{c}{\textbf{PWiseR}} & \multicolumn{1}{c}{\textbf{BCE}} & \multicolumn{1}{c}{\textbf{PWiseR}} &  & \multicolumn{1}{c}{}                                      & \multicolumn{1}{c}{\textbf{BCE}} & \multicolumn{1}{c}{\textbf{PWiseR}} & \multicolumn{1}{c}{\textbf{BCE}} & \multicolumn{1}{c}{\textbf{PWiseR}} \\ \cline{1-5} \cline{7-11} 
MMoE                                                     & \textbf{71.214}                & 71.158                            & 62.379                         & \textbf{62.716}                   &  & MMoE                                                      & 72.846                         & \textbf{73.069}                   & 64.257                         & \textbf{64.284}                   \\ \cline{1-5} \cline{7-11} 
PLE                                                      & 71.149                         & \textbf{71.175}                   & \textbf{62.365}                & 62.126                            &  & PLE                                                       & 72.883                         & \textbf{72.894}                   & 64.067                         & \textbf{64.369}                   \\ \cline{1-5} \cline{7-11} 
SharedBottom                                             & 70.883                         & \textbf{70.941}                   & \textbf{62.200}                & 61.937                            &  & SharedBottom                                              & \textbf{72.741}                 & 72.614                            & \textbf{63.941}                & 63.932                            \\ \cline{1-5} \cline{7-11} 
                                                         &                                  &                                     &                                  &                                     &  &                                                           &                                  &                                     &                                  &                                     \\
\multicolumn{1}{c}{\multirow{2}{*}{\textbf{Alibaba NL}}} & \multicolumn{2}{c}{\textbf{CTR}}                                       & \multicolumn{2}{c}{\textbf{CTCVR}}                                     &  & \multicolumn{1}{c}{\multirow{2}{*}{\textbf{Alibaba CCP}}} & \multicolumn{2}{c}{\textbf{CTR}}                                       & \multicolumn{2}{c}{\textbf{CTCVR}}                                     \\
\multicolumn{1}{c}{}                                     & \multicolumn{1}{c}{\textbf{BCE}} & \multicolumn{1}{c}{\textbf{PWiseR}} & \multicolumn{1}{c}{\textbf{BCE}} & \multicolumn{1}{c}{\textbf{PWiseR}} &  & \multicolumn{1}{c}{}                                      & \multicolumn{1}{c}{\textbf{BCE}} & \multicolumn{1}{c}{\textbf{PWiseR}} & \multicolumn{1}{c}{\textbf{BCE}} & \multicolumn{1}{c}{\textbf{PWiseR}} \\ \cline{1-5} \cline{7-11} 
MMoE                                                     & 72.890                         & \textbf{72.991}                   & 63.196                         & \textbf{63.336}                   &  & MMoE                                                      & 61.852                         & \textbf{61.859}                   & \textbf{40.700}                & 40.399                            \\ \cline{1-5} \cline{7-11} 
PLE                                                      & 72.642                         & \textbf{72.668}                   & 62.778                         & \textbf{62.856}                   &  & PLE                                                       & 61.481                         & \textbf{61.491}                   & 39.774                         & \textbf{40.082}                   \\ \cline{1-5} \cline{7-11} 
SharedBottom                                             & 72.924                         & \textbf{73.000}                   & 63.008                         & \textbf{63.266}                   &  & SharedBottom                                              & 61.539                         & \textbf{61.840}                   & 38.853                & \textbf{38.928}
                            \\ \cline{1-5} \cline{7-11} 
\end{tabular}}
\end{table*}

\textbf{Performance on STL models}:
Table \ref{STL} presents the results of STL CTR prediction using both BCE and PWiseR losses for four public datasets across two models: DNN and MaskNet.
In the DNN model, the proposed PWiseR loss improves performance across all datasets except for Alibaba NL. Similarly, for the MaskNet model, PWiseR enhances results in all datasets except for Alibaba US.

\begin{table}[H]
\centering
\caption{The table shows single task CTR prediction of both BCE and PWiseR losses for 4 public datasets.}
\label{STL}
\begin{tabular}{lllll}
\multirow{2}{*}{\textbf{Dataset}} & \multicolumn{2}{c}{\textbf{DNN}}      & \multicolumn{2}{c}{\textbf{MaskNet}}  \\ \cline{2-5} 
                         & \textbf{BCE}      & \textbf{PWiseR}   & \textbf{BCE}      & \textbf{PWiseR}   \\ \hline
Alibaba US               & 70.946          & \textbf{71.112} & \textbf{71.350} & 71.323           \\ \hline
Alibaba FR               & 72.865           & \textbf{73.015} & 72.931          & \textbf{72.943} \\ \hline
Alibaba NL               & \textbf{73.130} & 72.935          & 72.912          & \textbf{73.001} \\ \hline
Alibaba CCP              & 62.255          & \textbf{62.310} & 62.431          & \textbf{62.441} \\ \hline
\end{tabular}
\end{table}

\textbf{Performance on Industrial dataset}:
Table \ref{industrial} provides a concise overview of the MMoE and PLE models' performance on an industrial dataset, focusing on Click-Through Rate (CTR) and Click-Through Conversion Rate (CTCVR) metrics under both BCE and the proposed PWiseR losses. It's evident that transitioning to the PWiseR loss leads to improvements in the AUC results for both CTR and CTCVR across both models. Specifically, for MMoE, the CTR increases from 79.413 to 79.592, and the CTCVR improves from 69.974 to 70.153. Similarly, for PLE, the CTR rises from 80.664 to 80.775, and the CTCVR enhances from 71.017 to 71.139. These findings highlight the effectiveness of the PWiseR loss in enhancing the predictive performance of both MMoE and PLE models on industrial datasets, particularly in optimizing click-through rates and conversion rates in online advertising contexts.

\begin{table}[h]
\centering
\caption{The table shows the results of industrial dataset experiments in terms of AUC(\%) metric.}
\label{industrial}
\begin{tabular}{lllll}
\multicolumn{1}{c}{\multirow{2}{*}{\textbf{Industrial}}} & \multicolumn{2}{c}{\textbf{CTR}} & \multicolumn{2}{c}{\textbf{CTCVR}} \\ \cline{2-5} 
\multicolumn{1}{c}{}                                     & \textbf{BCE}       & \textbf{PWiseR}  & \textbf{BCE}         & \textbf{PWiseR}     \\ \hline
MMoE                                                     & 79.413                & \textbf{79.592}           & 69.974                  & \textbf{70.153}              \\ \hline
PLE                                                      & 80.664  & \textbf{80.775}      & 71.017     & \textbf{71.139}                       \\ \hline
\end{tabular}
\end{table}

\section{CONCLUSION}

This paper presents a novel approach for multi-task learning (MTL) in online advertising systems, specifically targeting the challenges in click-through rate (CTR) and conversion rate (CVR) prediction. Recognizing the inherent sequential relationship between CTR and CVR tasks, we introduced a task-specific PWiseR loss to address the issue of noisy data and improve prediction accuracy.

Our approach emphasizes the importance of distinguishing between samples with and without conversions. By leveraging the sequential pattern of user interactions (impression -> click -> conversion), the proposed PWiseR loss function encourages the model to prioritize samples with conversions, leading to more reliable and accurate predictions. This is achieved by penalizing the model for assigning lower scores to conversion samples compared to click-only samples and non-click samples.

Through extensive experiments on both public and industrial datasets, our proposed method consistently outperformed traditional binary cross-entropy (BCE) loss in terms of the AUC metric. This demonstrates the effectiveness of the PWiseR loss in enhancing model performance by reducing the impact of noisy data and improving the ranking accuracy of high-value samples.

\newpage
\bibliographystyle{unsrt}  
\bibliography{sample-base}

\begin{thebibliography}{10}

\bibitem{cvrattention}
Yumin Su, Liang Zhang, Quanyu Dai, Bo~Zhang, Jinyao Yan, Dan Wang, Yongjun Bao, Sulong Xu, Yang He, and Weipeng Yan.
\newblock An attention-based model for conversion rate prediction with delayed feedback via post-click calibration.
\newblock In {\em Proceedings of the Twenty-Ninth International Joint Conference on Artificial Intelligence}, IJCAI'20, 2021.

\bibitem{ctnocvr}
Dandan Zhang, Haotian Wu, Guanqi Zeng, Yao Yang, Weijiang Qiu, Yujie Chen, and Haoyuan Hu.
\newblock Ctnocvr: A novelty auxiliary task making the lower-ctr-higher-cvr upper.
\newblock In {\em Proceedings of the 45th International ACM SIGIR Conference on Research and Development in Information Retrieval}, SIGIR '22, page 2272–2276, New York, NY, USA, 2022. Association for Computing Machinery.

\bibitem{star}
Jianqiang Huang, Ke~Hu, Qingtao Tang, Mingjian Chen, Yi~Qi, Jia Cheng, and Jun Lei.
\newblock Deep position-wise interaction network for ctr prediction.
\newblock In {\em Proceedings of the 44th International ACM SIGIR Conference on Research and Development in Information Retrieval}, SIGIR '21, page 1885–1889, New York, NY, USA, 2021. Association for Computing Machinery.

\bibitem{3MN}
Yifei Zhang, Hua Hua, Hui Guo, Shuangyang Wang, Chongyu Zhong, and Shijie Zhang.
\newblock 3mn: Three meta networks for multi-scenario and multi-task learning in online advertising recommender systems.
\newblock In {\em Proceedings of the 32nd ACM International Conference on Information and Knowledge Management}, CIKM '23, page 4945–4951, New York, NY, USA, 2023. Association for Computing Machinery.

\bibitem{caruana1997multitask}
Rich Caruana.
\newblock Multitask learning.
\newblock {\em Machine learning}, 28:41--75, 1997.

\bibitem{esmm}
Xiao Ma, Liqin Zhao, Guan Huang, Zhi Wang, Zelin Hu, Xiaoqiang Zhu, and Kun Gai.
\newblock Entire space multi-task model: An effective approach for estimating post-click conversion rate.
\newblock In {\em The 41st International ACM SIGIR Conference on Research \& Development in Information Retrieval}, SIGIR '18, page 1137–1140, New York, NY, USA, 2018. Association for Computing Machinery.

\bibitem{mmoe}
Jiaqi Ma, Zhe Zhao, Xinyang Yi, Jilin Chen, Lichan Hong, and Ed~H Chi.
\newblock Modeling task relationships in multi-task learning with multi-gate mixture-of-experts.
\newblock In {\em Proceedings of the 24th ACM SIGKDD international conference on knowledge discovery \& data mining}, pages 1930--1939, 2018.

\bibitem{ple}
Hongyan Tang, Junning Liu, Ming Zhao, and Xudong Gong.
\newblock Progressive layered extraction (ple): A novel multi-task learning (mtl) model for personalized recommendations.
\newblock In {\em Proceedings of the 14th ACM Conference on Recommender Systems}, pages 269--278, 2020.

\bibitem{hinet}
Jie Zhou, Xianshuai Cao, Wenhao Li, Lin Bo, Kun Zhang, Chuan Luo, and Qian Yu.
\newblock Hinet: Novel multi-scenario multi-task learning with hierarchical information extraction.
\newblock In {\em 2023 IEEE 39th International Conference on Data Engineering (ICDE)}, pages 2969--2975, 2023.

\bibitem{causalint}
Yichao Wang, Huifeng Guo, Bo~Chen, Weiwen Liu, Zhirong Liu, Qi~Zhang, Zhicheng He, Hongkun Zheng, Weiwei Yao, Muyu Zhang, Zhenhua Dong, and Ruiming Tang.
\newblock Causalint: Causal inspired intervention for multi-scenario recommendation.
\newblock In {\em Proceedings of the 28th ACM SIGKDD Conference on Knowledge Discovery and Data Mining}, KDD '22, page 4090–4099, New York, NY, USA, 2022. Association for Computing Machinery.

\bibitem{ip}
Conor O'Brien, Kin~Sum Liu, James Neufeld, Rafael Barreto, and Jonathan~J Hunt.
\newblock An analysis of entire space multi-task models for post-click conversion prediction.
\newblock In {\em Proceedings of the 15th ACM Conference on Recommender Systems}, RecSys '21, page 613–619, New York, NY, USA, 2021. Association for Computing Machinery.

\bibitem{escd}
Yunfeng Zhao, Xu~Yan, Xiaoqiang Gui, Shuguang Han, Xiang-Rong Sheng, Guoxian Yu, Jufeng Chen, Zhao Xu, and Bo~Zheng.
\newblock Entire space cascade delayed feedback modeling for effective conversion rate prediction.
\newblock In {\em Proceedings of the 32nd ACM International Conference on Information and Knowledge Management}, CIKM '23, page 4981–4987, New York, NY, USA, 2023. Association for Computing Machinery.

\bibitem{ruder2017overview}
Sebastian Ruder.
\newblock An overview of multi-task learning in deep neural networks.
\newblock {\em arXiv preprint arXiv:1706.05098}, 2017.

\bibitem{xi2021modeling}
Dongbo Xi, Zhen Chen, Peng Yan, Yinger Zhang, Yongchun Zhu, Fuzhen Zhuang, and Yu~Chen.
\newblock Modeling the sequential dependence among audience multi-step conversions with multi-task learning in targeted display advertising.
\newblock In {\em Proceedings of the 27th ACM SIGKDD Conference on Knowledge Discovery \& Data Mining}, pages 3745--3755, 2021.

\bibitem{caruana1993multitask}
R~Caruana.
\newblock Multitask learning: A knowledge-based source of inductive bias1.
\newblock In {\em Proceedings of the Tenth International Conference on Machine Learning}, pages 41--48. Citeseer, 1993.

\bibitem{covington2016deep}
Paul Covington, Jay Adams, and Emre Sargin.
\newblock Deep neural networks for youtube recommendations.
\newblock In {\em Proceedings of the 10th ACM conference on recommender systems}, pages 191--198, 2016.

\bibitem{moe}
Robert~A Jacobs, Michael~I Jordan, Steven~J Nowlan, and Geoffrey~E Hinton.
\newblock Adaptive mixtures of local experts.
\newblock {\em Neural computation}, 3(1):79--87, 1991.

\bibitem{esdf}
Yanshi Wang, Jie Zhang, Qing Da, and Anxiang Zeng.
\newblock Delayed feedback modeling for the entire space conversion rate prediction.
\newblock 2020.

\bibitem{mtms}
Shulong Tan, Meifang Li, Weijie Zhao, Yandan Zheng, Xin Pei, and Ping Li.
\newblock Multi-task and multi-scene unified ranking model for online advertising, December 2021.

\bibitem{mmn}
Wentao Ouyang, Xiuwu Zhang, Chaofeng Guo, Shukui Ren, Yupei Sui, Kun Zhang, Jinmei Luo, Yunfeng Chen, Dongbo Xu, Xiangzheng Liu, and Yanlong Du.
\newblock Masked multi-domain network: Multi-type and multi-scenario conversion rate prediction with a single model.
\newblock {\em arXiv}, 2024.

\bibitem{mebs}
Zhen Gong, Lvyin Niu, Yang Zhao, Miao Xu, Zhenzhe Zheng, Haoqi Zhang, Zhilin Zhang, Fan Wu, Rongquan Bai, Chuan Yu, Jian Xu, and Bo~Zheng.
\newblock Mebs: Multi-task end-to-end bid shading for multi-slot display advertising.
\newblock 2024.

\bibitem{dupn}
Yabo Ni, Dan Ou, Shichen Liu, Xiang Li, Wenwu Ou, Anxiang Zeng, and Luo Si.
\newblock Perceive your users in depth: Learning universal user representations from multiple e-commerce tasks.
\newblock 2018.

\bibitem{deepfm}
Huifeng Guo, Ruiming Tang, Yunming Ye, Zhenguo Li, and Xiuqiang He.
\newblock Deepfm: a factorization-machine based neural network for ctr prediction.
\newblock In {\em Proceedings of the 26th International Joint Conference on Artificial Intelligence}, IJCAI'17, page 1725–1731. AAAI Press, 2017.

\bibitem{mmbattn}
Hasan Saribas, Cagri Yesil, Serdarcan Dilbaz, and Halit Orenbas.
\newblock Mmbattn: Max-mean and bit-wise attention for ctr prediction, 2023.

\bibitem{MaskNet}
Zhiqiang Wang, Qingyun She, and Junlin Zhang.
\newblock Masknet: Introducing feature-wise multiplication to ctr ranking models by instance-guided mask, 2021.

\bibitem{stec}
Serdarcan Dilbaz and Hasan Saribas.
\newblock Stec: See-through transformer-based encoder for ctr prediction, 2023.

\bibitem{ma2018entire}
Xiao Ma, Liqin Zhao, Guan Huang, Zhi Wang, Zelin Hu, Xiaoqiang Zhu, and Kun Gai.
\newblock Entire space multi-task model: An effective approach for estimating post-click conversion rate, 2018.

\bibitem{peng2020improving}
pengcheng Li, Runze Li, Qing Da, An-Xiang Zeng, and Lijun Zhang.
\newblock Improving multi-scenario learning to rank in e-commerce by exploiting task relationships in the label space.
\newblock In {\em proceedings of the 28th {ACM} International Conference on Information and Knowledge Management, {CIKM} 2020, Virtual Event, Ireland, October 19- 23,2019}, New York,NY,USA, 2020. {ACM}.

\bibitem{zhu2021learning}
Yongchun Zhu, Yudan Liu, Ruobing Xie, Fuzhen Zhuang, Xiaobo Hao, Kaikai Ge, Xu~Zhang, Leyu Lin, and Juan Cao.
\newblock Learning to expand audience via meta hybrid experts and critics for recommendation and advertising.
\newblock In {\em Proceedings of the 27th ACM SIGKDD Conference on Knowledge Discovery \& Data Mining}, pages 4005--4013, 2021.

\bibitem{ZhuLYZH21}
Jieming Zhu, Jinyang Liu, Shuai Yang, Qi~Zhang, and Xiuqiang He.
\newblock Open benchmarking for click-through rate prediction.
\newblock In Gianluca Demartini, Guido Zuccon, J.~Shane Culpepper, Zi~Huang, and Hanghang Tong, editors, {\em {CIKM} '21: The 30th {ACM} International Conference on Information and Knowledge Management, Virtual Event, Queensland, Australia, November 1 - 5, 2021}, pages 2759--2769. {ACM}, 2021.

\bibitem{ZhuDSMLCXZ22}
Jieming Zhu, Quanyu Dai, Liangcai Su, Rong Ma, Jinyang Liu, Guohao Cai, Xi~Xiao, and Rui Zhang.
\newblock {BARS:} towards open benchmarking for recommender systems.
\newblock In Enrique Amig{\'{o}}, Pablo Castells, Julio Gonzalo, Ben Carterette, J.~Shane Culpepper, and Gabriella Kazai, editors, {\em {SIGIR} '22: The 45th International {ACM} {SIGIR} Conference on Research and Development in Information Retrieval, Madrid, Spain, July 11 - 15, 2022}, pages 2912--2923. {ACM}, 2022.

\end{thebibliography}

\appendix

\end{document}